# Single Artificial Atom SASER

Shtefan V. Sanduleanu[1,*], Peter Yu. Shlykov[2,1,*], Alexei N. Bolgar[1], Daria A. Kalacheva[1], Julia I. Zotova[1], Gleb P. Fedorov[1], Viktor B. Lubsanov[1], Alexei Yu. Dmitriev[1], Evgenia S. Alekseeva[1], Oleg V. Astafiev[2,1]

[1]Moscow Center for Advanced Studies, Moscow 123592, Russia
[2]Center for Engineering Physics, Moscow 121205, Russia

**Abstract**

Lasing – an effect of orthodox quantum mechanics – was discovered in 1955 and recognized by the Nobel Prize in 1964 due to its fundamentality[1,2]. Nowadays, lasers and masers routinely work with electromagnetic waves and consist of a resonator with an active medium – usually a system of atoms with population inversion mechanism. Amazingly, quantum mechanics remains valid even when electromagnetic waves are replaced by vibrations of a crystal lattice[3-6], and, therefore, photons by phonons, even though are not fundamental particles. By implementing acoustic resonators coupled to an atom with a mechanism of population inversion, the lasing effect in sound can be achieved.

In this paper, we demonstrate the single artificial atom SASER (Sound Amplification by Stimulated Emission of Radiation) action by utilizing a surface acoustic wave (SAW) resonator on quartz[7] coupled to a deliberately designed superconducting three-level quantum system (artificial atom), in which population inversion is realized[8]. The SASER operates in the ultrasound range at a frequency about 3 GHz. Acoustic-to-electric signals are converted via piezo-electric effect and the circuit elements; an artificial atom and input/outputs are coupled via the acoustic waves. We observe amplification of the waves and their strong self-emission with a significant narrowing of the linewidth. The phonon number generated in the system exceeds 90.

**Introduction**

Superconducting quantum circuits have emerged as a highly versatile platform for quantum optics. They enable the observation of many fundamental phenomena such as the strong coupling of a single photon to a superconducting qubit[9], resonance fluorescence from an artificial atom[10], and many others[11-13]. This success in circuit quantum electrodynamics (cQED) is based on the tuneable, strong light-matter interactions, which these systems provide[9,10]. Remarkably, the formalism of quantum mechanics remains applicable when electromagnetic fields and photons are replaced by vibrations of a crystal lattice and phonons.

* with equal contribution, shtefan.sanduleanu@gmail.com, shlykov.piu@gmail.com

The field of circuit quantum acoustodynamics (cQAD) utilizes the piezoelectric effect in materials like quartz to interface superconducting circuits with surface acoustic waves (SAWs)[14,5] or high-quality SAW resonators[7]. SAW resonators, formed by Bragg mirrors made of metallic stripes on top of quartz, can reach frequencies up to a few gigahertz and quality factors up to $10^5$. Recent experimental demonstrations include the coupling of superconducting artificial atoms to mechanical degrees of freedom[15-17], quantum control of propagating phonons and phonon-mediated state transfer[14], highlighting the rich potential of phononic quantum effects.

A conventional laser is a device based on the fundamental quantum mechanical principle of stimulated emission from a gain medium, represented by an ensemble of atoms subject to population inversion confined within an electromagnetic cavity. In particular, the lasing effect with a single artificial atom has previously been demonstrated in superconducting systems[18,19]. By combining an acoustic resonator with an artificial atom driven to population inversion, the acoustic analogue to a laser – the SASER – can be realized. Recently, electrically induced acoustic laser emission stated in a mechanism similar to that of solid-state diode lasers[20]. However, the classical lasing mechanism in sound with population inversion in a quantized system (atom) yet has to be demonstrated.

Here, we demonstrate the SASER (Sound Amplification by Stimulated Emission of Radiation) operation by coupling a SAW resonator on quartz[7] to a specifically designed superconducting three-level artificial atom. The resonant frequency of the acoustic resonator, and consequently the SASER operation, is 3.21 GHz. The resulting device shows clear lasing signatures: signal reaching amplification by 8 times and significant narrowing of the emission linewidth by a factor of 9.

**Device description**

The SASER operation scheme is illustrated in Fig. 1a: a three-level artificial atom is coupled to an acoustic resonator. Our device relies on the following elements and achievements: (i) a SAW phononic crystal with resonant acoustic fields confined in a small 2D volume in both *x* and *y* directions; (ii) a superconducting three-level artificial atom implemented in a flux-qubit geometry (a loop interrupted by a series of Josephson junctions) to enable population inversion; (iii) physically strong coupling of the artificial atom to the acoustic field of the SAW resonator; (iv) additional (strong) coupling of the artificial atom to an electric field of a transmission line – 1D free space.

Figures 1b-c show the device schematics together with its micrograph. The device is fabricated on top of a quartz substrate. The artificial atom (blue) is coupled both to the acoustic resonator (green) and to an open transmission line (magenta). The resonator is bounded by two Bragg mirrors (green structure) consisting of aluminium stripes with a period of 0.959 μm. The inner structure between the mirrors (red and blue) is formed with Interdigital Transducers (IDTs) with a period of pairs of aluminum stripes

(ground signal) of 0.950 μm[7]. The role of IDTs is twofold: to form a phononic crystal with resonant modes[16] and to convert electric fields into/from acoustic fields. The integrated design combining the phononic crystal with Bragg mirrors suppresses diffraction losses due to total internal reflection at the edges of the structure[7]. Furthermore, the reduced transversal sizes help to concentrate both electromagnetic energy and mechanical energy, thereby enhancing electromechanical coupling efficiency.

The artificial atom (blue colour in Fig. 1) consists of a loop with Josephson junctions, fabricated via the standard two-angle aluminium shadow deposition. It incorporates a small SQUID alpha-loop[21]. The area ratio between the main flux loop and SQUID loop is 12:1. Such a geometry enables precise energy tuning through external magnetic flux bias, which is controlled with magnetic field from a superconducting coil. The estimated Josephson energy of the large junctions is $E_J/h = 74$ GHz ($h$ is the Plank's constant), Josephson energy of the alpha-loop is $E_{J\alpha}/h = 56$ GHz and charging energy is $E_C/h = 0.32$ GHz.

The 'atom' is coupled to the resonator acoustic field through the central IDT with coupling capacitance $C_{ar} = 9.5$ fF. With an IDT length extending over several wavelengths, the system operates in the giant-atom regime[3,22]. Two adjacent IDTs with capacitances $C_{in} = C_{out} = 10$ fF convert microwave from the input line to the acoustic field and the acoustic field to the output microwave line. To measure acoustic fields and the transmission through the resonator in/out coplanar lines are used (red colour). Importantly, the signal transduction efficiency via the acoustic field (electrical→mechanical→electrical) exceeds direct capacitive in-out port (electromagnetic) coupling by five orders of magnitude, therefore the ports are coupled effectively solely via the surface acoustic waves.

In our circuit we implement an additional transmission line (magenta in Fig. 1), so that the artificial atom is strongly coupled to that line (1D open space) with an IDT of capacitance $C_{at} = 32$ fF (Fig. 1b, c). This allows excitation of the artificial atom by propagating microwaves and relaxation via photon emission into the line with corresponding transition frequencies $\omega_{eg}$, $\omega_{fg}$ and $\omega_{fe}$ of $|e\rangle \leftrightarrow |g\rangle$, $|f\rangle \leftrightarrow |g\rangle$ and $|f\rangle \leftrightarrow |e\rangle$ transitions. The choice of the flux-qubit-geometry is deliberate: unlike capacitively shunted charge qubits, the flux geometry permits the $|g\rangle \rightarrow |f\rangle$ transition, enabling system pumping via external fields[8]. This transition is forbidden in transmon qubits.

## Results

The spectroscopic characterization of the artificial atom was performed via single-tone notch-type transmission measurements through the coplanar waveguide (magenta in Fig.1). We acquired spectroscopy data by sweeping the magnetic flux threading the superconducting loops of the alpha-loop flux qubit. The broadband spectra of the atomic transitions are shown in Fig. 2. In order to enable interaction between the atom and the resonator, the lowest atomic transition ($|g\rangle \leftrightarrow |e\rangle$) is brought into resonance with the SAW resonator. Key features of the flux qubit are visible in the spectroscopy data (Fig. 2), including a minimum in the $|e\rangle \rightarrow |g\rangle$ relaxation rate and the vanishing of the $|f\rangle \rightarrow |g\rangle$ transition line at the lowest ('sweet spot') energy.

We deduce the atom's parameters near the resonance. The total relaxation rates are $\Gamma_{eg}/2\pi = 35$ MHz, $\Gamma_{fg}/2\pi = 6$ MHz and $\Gamma_{fe}/2\pi = 100$ MHz. The relations $\Gamma_{fe} \gg \Gamma_{fg}$ and $\Gamma_{fe} > \Gamma_{eg}$ satisfy the conditions for population inversion. The fundamental mode of the resonator with frequency $\omega_r/2\pi = 3.21$ GHz shows decay rates of $\kappa_0/2\pi = 134$ kHz (single-phonon) and $\kappa_m/2\pi = 94$ kHz (multi-phonon regime), corresponding to quality factors of $Q_0 = 24\times10^3$ and $Q_m = 34\times10^3$. From the avoided crossing observed by tuning the qubit frequency, we determined a coupling strength between the resonator and the atom ($|g\rangle \leftrightarrow |e\rangle$) of $g/2\pi = 11$ MHz.

The device was characterised near the flux bias point, where vacuum Rabi splitting is observed in the resonator's transmission spectrum (Fig. 3a). The introduction of an additional microwave pump tone at frequency $\omega_{fg}$ to the transmission line alters the response dramatically: the avoided crossing vanishes and is replaced by "hot-spots", where the transmission exceeds unity (Fig. 3b).

We performed detailed spectroscopy at one of these hot-spots. Transmission measurements with reduced probe tone power ($5 \times 10^{-18}$ W) reveal enhanced gain and a strong linewidth narrowing (Fig. 3c). In Fig. 3d the transmission band narrows by a factor of 9, and the signal shows amplification by more than 8 times in amplitude.

Next, we measure the emission from the SASER using a spectrum analyser. The power emitted from the resonator was measured as a function of frequency and external flux (Fig. 4a). Both the emission and the amplified transmission (Fig. 3c) peaks occur at the same frequency, which is detuned from the resonator frequency. When the atom is in the resonance and is pumped at a frequency of the $|g\rangle \leftrightarrow |f\rangle$ transition by $P = 0.1$ pW a strong narrow emission peak is observed (Fig. 4b) on a top of the flat amplifier noise background. Figure 4b compares the emission spectrum at a fixed flux bias (red) to the squared magnitude of the transmission (blue) ($|t|^2$) measured using a weak ($5 \times 10^{-18}$ W) probe tone. The full-width at half-maximum (FWHM) of the emission peak is nine times narrower than that of the fundamental resonator mode, confirming the onset of a lasing. The measured emission peak with subtracted background is best fitted by a Voigt profile, indicating the presence of Gaussian noise processes.

We estimate the phonon population ($N_{pn}$) in the lasing state from the measured output power and the calibrated gain of the output line. The maximum phonon number in our experiment is approximately 90. Figure 4d shows the relation of the linewidth (FWHM) $\Delta\omega$ from emission spectroscopy measurements (Fig. 4a) on the phonon number. The red curve exemplifies the inverse square root dependence $\Delta\omega = 2\kappa_m/\sqrt{2N_{pn}}$ as an eye guide. Note that the theoretical limit due to quantum noise (Schallow-Townes formula) is $\frac{\kappa}{2N_{pn}}$; however, this limit is challenging to realise experimentally due to various line-broadening mechanisms.

**Calculations**

Our three-level system driven by an external field with an amplitude $\Omega$ between the ground and the second excited states ($|g\rangle \leftrightarrow |f\rangle$) coupled to the single-mode resonator is described in the rotating wave approximation by the Hamiltonian

$$H = \hbar\delta\omega_{ge}\sigma_{ee} + \frac{\hbar\Omega}{2}(\sigma_{gf} + \sigma_{fg}) + g(\sigma_{ge}b + \sigma_{eg}b^\dagger), \quad (1)$$

where the atomic projection/transition operators are $\sigma_{ij} = |i\rangle\langle j|$ with $i$ and $j$ taking ground ($g$), excited ($e$) or second excited ($f$) states; $b^\dagger$ and $b$ are the phonon creation and annihilation operators in the SAW resonator; $\delta\omega_{ge} = \omega_{ge} - \omega_r$ is the detuning between the resonator and the lowest atomic transition. By solving the master equation, incorporating Lindblad terms to account for all damping mechanisms, one can obtain the numerical solutions for both the system dynamics and the steady state. However, for large phonon numbers, exact diagonalization becomes computationally intractable due to the prohibitively large Hilbert space of the system.

Nevertheless, the system dynamics can be modelled using semiclassical equations for the expectation values of the atomic and SAW resonator operators. Under the assumption of factorization of mean values of the atom-resonator correlators, and employing the quantum regression theorem, we obtain a set of equations describing the system dynamics, emission spectra, and steady-state properties. A representative calculation is shown in Fig. 4c, which reproduces the experimental data represented in Fig. 4a.

In the steady state, the system satisfies the following rate equation

$$\Gamma_{fe}\langle\sigma_{ff}\rangle - \Gamma_{eg}\langle\sigma_{ee}\rangle = \kappa N_{pn}, \quad (2)$$

which represents equilibration between excitations and decay in the atom-resonator system. This allows the mean phonon number to be calculated numerically. However, we also provide an intuitive picture that explains its behaviour. We consider driving conditions, corresponding to the maximal possible $N_{pn}$ at the resonance ($\delta\omega_{ge} = 0$). The following conditions to be fulfilled in the simplified model: (i) atom-resonator coupling and the drive are much stronger than any incoherent processes: $\Omega \gg (\Gamma_{fe}, \Gamma_{fg})$, $g\sqrt{N_{pn}} \gg (\Gamma_{eg}, \kappa)$; (ii) the strong drive of the system $\Omega$ and the coupling $g\sqrt{N_{pn}}$ saturate the transitions: $\langle\sigma_{gg}\rangle \approx \langle\sigma_{ff}\rangle$ and $\langle\sigma_{gg}\rangle \approx \langle\sigma_{ee}\rangle$ correspondingly, resulting in roughly equal populations $\langle\sigma_{gg}\rangle \approx \langle\sigma_{ee}\rangle \approx \langle\sigma_{ff}\rangle$ and, therefore, are about 1/3 each; (iii) the optimal pumping condition occurs near $\Omega/2 \approx g\sqrt{N_{pn}}$, so that both effective fields are equilibrated.

A consequence of assumption (ii) is that the maximal phonon number is given by:

$$N_{pn}^{max} \approx \frac{\Gamma_{fe}-\Gamma_{eg}}{3\kappa}. \quad (3)$$

Substituting the system parameters, we find $N_{pn}^{max} \approx 200$, which is in good agreement with the more accurate numerical results. However, this model doesn't take into account coupling for $|e\rangle \leftrightarrow |f\rangle$ transition to the acoustic resonator, which modifies the phonon number. For example it causes Purcell effect for the resonator and enlarges its' decay rate proportional to $\kappa_{Purc} \sim \frac{g_{fe}^2 \Gamma_{fe}}{\Delta^2}$, where $g_{fe}$ is a coupling strength for $|e\rangle \leftrightarrow |f\rangle$ transition to the resonator and $\Delta$ is a detuning of the $|e\rangle \leftrightarrow |f\rangle$ transition frequency from the resonator $\Delta = \omega_{ef} - \omega_r$, which is equal to the atom's anharmonicity in resonance. This approximately doubles effective $\kappa$ in Eq. (3) and reduces twice the number of phonons. More accurate numerical estimate accounting anharmonicity gives the maximum phonon number to be $N_{pn}^{max} \approx 90$.

In the simplified model the generation threshold for such a system occurs approximately when population inversion is achieved in the absence of the resonator[7], $\Omega = \sqrt{\Gamma_{fe}\Gamma_{eg}}$. The optimal pump power determined from an independent experiment is $\Omega/2\pi = 270$ MHz, agreeing well with the numerically calculated value of $\Omega/2\pi = 290$ MHz.

**Summary**
We have demonstrated the SASER action in a SAW resonator coupled to a superconducting artificial atom. The lasing effect takes place in spite of replacing electromagnetic by acoustic waves and photons by phonons. This fundamental result shows that lasing is universal phenomenon and can be extended to artificial atoms and phononic systems. We expect that such a device could be used as a source and an amplifier of acoustic waves.

## Author contributions

O.V.A. - proposed the device concept and planned the experiment. S.V.S., P.I.S. and A.N.B. made the experiments with an important contribution from G.P.F., J.I.Z., A.Y.D., O.V.A. Samples were designed by A.N.B. and S.V.S., A.N.B. and J.I.Z. - developed fabrication technology for acoustic devices. A.N.B., D.A.K., V.B.L., A.Y.D. and E.S.A. - fabricated samples. S.V.S., P.I.S., A.N.B. and O.V.A. analysed data, prepared figures, provided theory and simulations of the experiment. P.I.S., O.V.A. and S.V.S. wrote the manuscript.

## Competing interests

The authors declare no competing interests.

## Figures

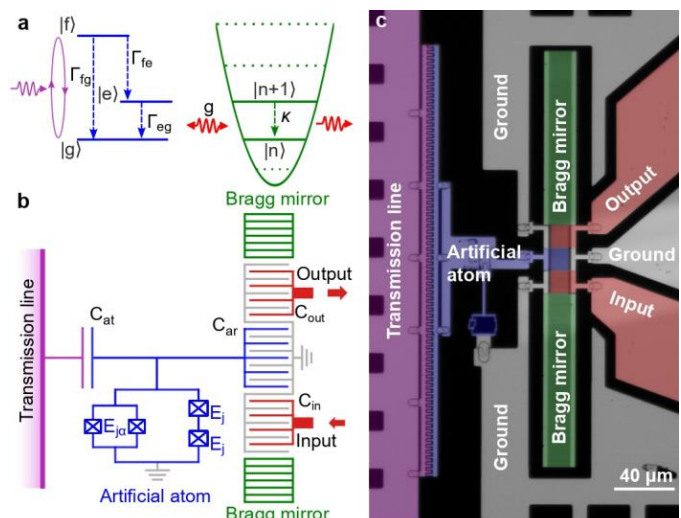

**Figure 1 | Single artificial-atom SASER and sasing mechanism. a**, Energy diagram of the system, describing the lasing process. **b**, Conceptual scheme of the experimental sample. Artificial atom is represented by alpha-loop SQUID flux qubit for better tuning of atomic properties via magnetic flux. Mechanical resonator is implemented as a phononic SAW crystal formed with IDTs and bounded by Bragg mirrors. **c**, Microphotograph of the device with false colours.

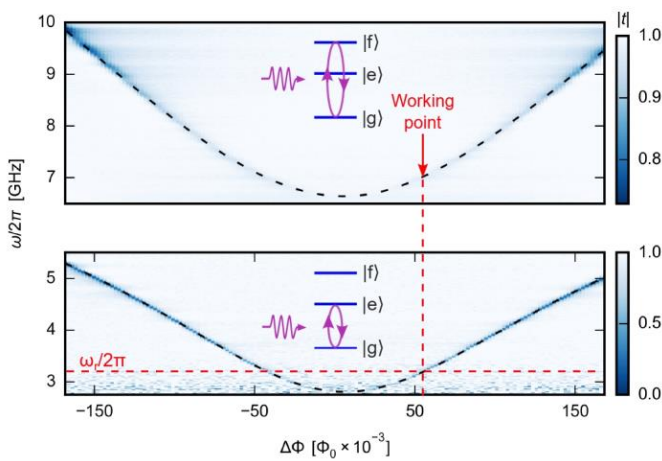

**Figure 2 | Transmission spectroscopy of the artificial atom through the transmission line.** The lower panel shows the transmission spectroscopy of the $|g\rangle \leftrightarrow |e\rangle$ flux qubit transition from the magnetic flux through the atom and the upper panel shows transmission spectroscopy of the $|g\rangle \leftrightarrow |f\rangle$ flux qubit transition from magnetic flux. The red horizontal striped line represents frequency of the acoustic resonator mode. The red vertical arrow and vertical red line show the working point, where atom and resonator are in resonance. Sketches represent the spectroscopy process.

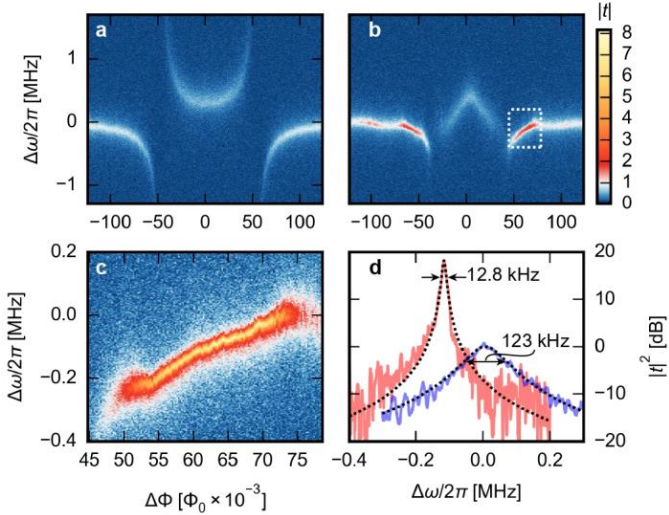

**Figure 3 | SASER transmission spectroscopy through the resonator between Input and Output lines. a**, Transmission spectroscopy of the acoustic mode at the avoided crossing between $|e\rangle \leftrightarrow |g\rangle$ flux qubit transition and acoustic mode. **b**, The same transmission spectroscopy of the acoustic resonator with the second tone pumping the artificial atom at frequency of $|g\rangle \leftrightarrow |f\rangle$ transition with power of 0.1 pW, highlighted region corresponds to the hot-spot (amplification), where lasing lakes place. **c**, Transmission spectroscopy of the resonator at the zoomed region with weak probe tone with atom pump power of 0.2 pW. **d**, Comparison of the transmitted signal through the resonator when the pump of the atom is on (red) and off (blue).

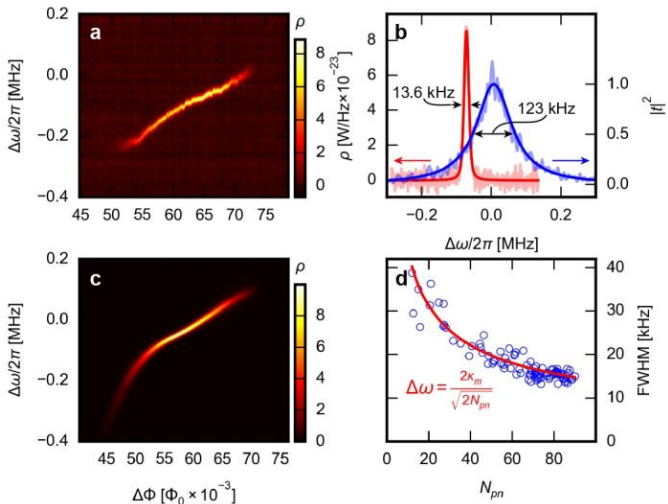

**Figure 4 | Emission from the saser at the Output. a**, Emission as a function of the magnetic flux at the zoomed region Fig. 3b with the same pumping power as in Fig. 3c. **b**, Comparison of radiation spectrum (red) and square of the transmission parameter $|t|^2$ (blue). **c**, Simulation of the SASER operation Fig. 4a with parameters, extracted from the experiment. **d**, Dependence of FWHM from the estimated number of phonons in the sasing state. The red curve is inverted square-root dependence.

# Supplementary Information

## Measurement Setup

All measurements were performed within a dilution refrigerator operating at a base temperature T ~ 20 mK. A schematic of the experimental setup and electronic circuitry are presented in Fig. S1. The sample is placed on the plate with a mixing chamber and connected with coaxial cables to the control and readout devices at room temperature. The atom is measured and pumped in a notch configuration via the microwave line and additionally controlled via magnetic flux generated with a superconducting coil. The resonator is also coupled to the readout line. A low-temperature switch is employed to determine the external quality factor of the acoustic resonator. Output signals pass through cryogenic isolators and are amplified by high electron mobility transistor (HEMT) amplifiers at 4K plate, followed by further amplification at room temperature. All transmission measurements are performed using a vector network analyser (VNA). A superconducting solenoid, driven by a DC current source, applies a flux bias to the artificial atom in to control its' energy level splitting. To additionally pump the artificial atom, we introduce an external Microwave source, coupled to the microwave line via a directional coupler. Emission power spectroscopy is performed simultaneously using a spectrum analyser with the VNA source disabled.

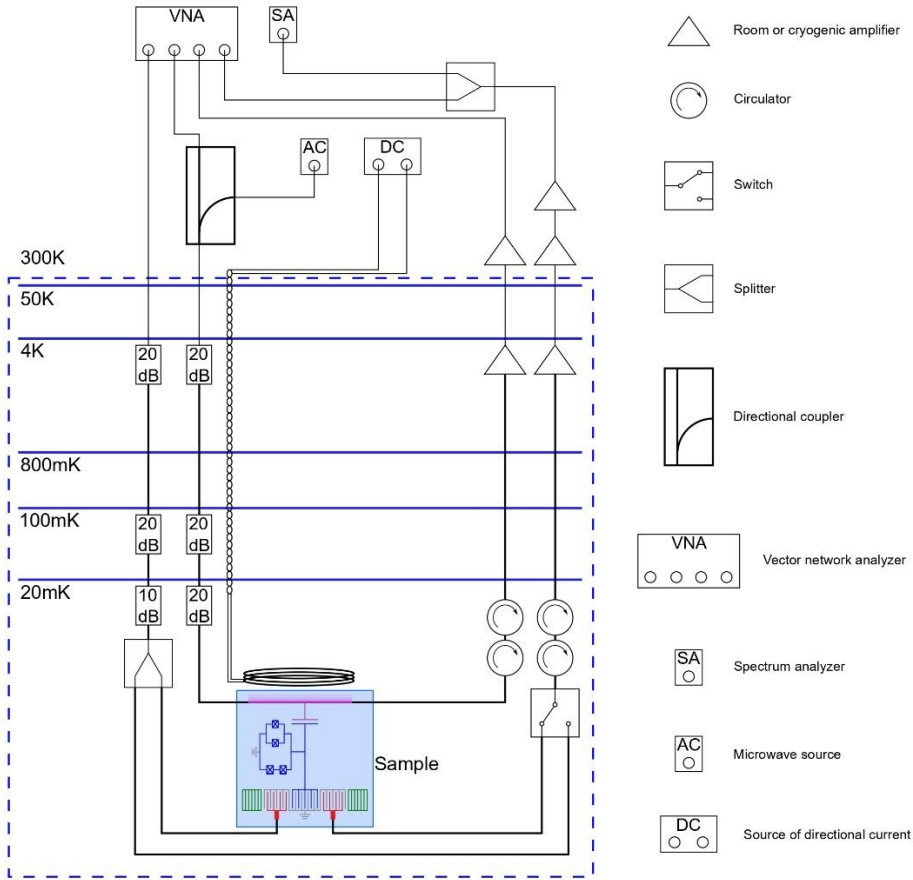

**Figure S1 |** Experimental setup and wiring.

## Artificial Atom

The spectroscopic characterization of the artificial atom was performed via single-tone notch-type transmission measurements. We acquired spectroscopy data by changing the magnetic flux threading the superconducting loops of the alpha-loop flux qubit. By biasing the flux near its lower sweet spot, the atom was brought into resonance with the acoustic resonator.

Figures S2a and S2d show the spectroscopy of the |g⟩ ↔ |e⟩ (ge) and |g⟩ ↔ |f⟩ (gf) transitions, respectively. The spectroscopy of the |e⟩ ↔ |f⟩ (ef) transition, presented in Fig. S2c, was measured by applying an additional microwave pump tone resonant with the gf transition. The difference between the fitted transition frequencies of the ge and gf transitions (from Fig. S2a and d) corresponds to the ef transition frequency measured in Fig. S2c across the measured flux range, confirming system integrity.

The decay rates for ge transition were extracted by fitting the notch-port transmission data with following formula: $S_{21}(f) = 1 - \frac{\Gamma_1}{2\Gamma_2} \times \frac{1+i(f-f_0)/\Gamma_2}{1+(f-f_0)^2/\Gamma_2^2 + \Omega^2/(\Gamma_1\Gamma_2)}$, where $\Gamma_1$ is a decoherence relaxation rate to the microwave line, $\Gamma_2$ is a total dephasing rate, $f_0$ is a transition frequency and $\Omega$ is a pumping rate. From the fit of the ge transition with pumping rate $\Omega$ = 13 MHz (Fig. S1a) we estimate full relaxation rate as $\Gamma_{eg}$ ≈ 35 MHz.

For the gf transition spectroscopy (Fig. S2d) we use formula derived from Master equations for three-level atom with probe tone at gf transition $S_{21}(f) = 1 - \frac{\Gamma_1}{2\Gamma_2} \times \frac{1}{1-2i(f-f_0)/\Gamma_2 + \frac{\Omega^2(2+\Gamma_{fe}/\Gamma_{eg})}{\Gamma_2(\Gamma_2+2i(f-f_0))}}$, yielding $\Gamma_1 = \Gamma_{fg}$ and $\Gamma_2 = \Gamma_{fe} + \Gamma_{fg}$. Recalculating Rabi frequency from attenuation of the line and probe tone power $\Omega$ = 30 MHz we get $\Gamma_{fg}$ = 6 MHz and $\Gamma_{fe}$ ≈ 100 MHz.

To corroborate our experimental results, we simulated the flux dependence of the transition frequencies using QuTiP. The results, overlaid as striped lines in Fig. S2a, c, d, show good agreement with the data. From these simulations, we extracted the Josephson energies of the junctions and the decay rates into the microwave line. The atom's capacitances were fixed based on independent finite-element simulations of the device geometry to simplify the parameter space. The relaxation rates calculated from the final simulated Hamiltonian are in good agreement with those obtained from the spectroscopic fits.

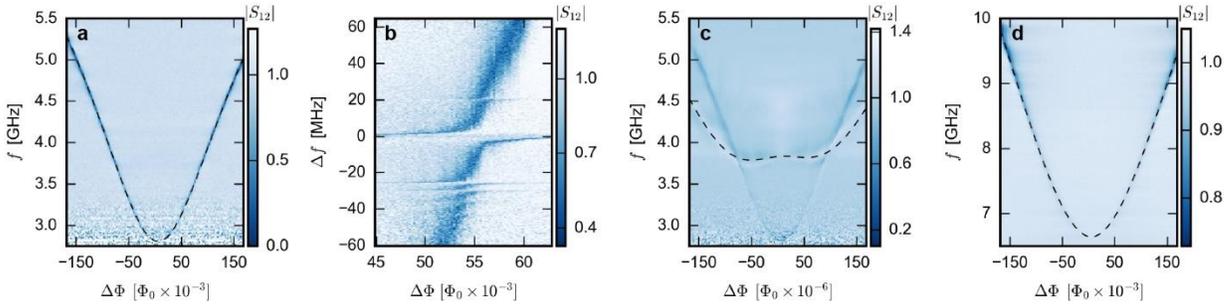

**Figure S2 | Notch-type transmission spectroscopy of the artificial atom. a**, Spectroscopy measurements of the |g⟩ ↔ |e⟩ transition. **b**, Spectroscopy measurements of the |g⟩ ↔ |e⟩ transition close to resonance of the acoustic resonator. Several resonator modes with different coupling strength are revealed. The fundamental mode, corresponding to the strongest coupling of g = 11 MHz is at f = 3.21 GHz. **c**, Spectroscopy measurements of the |e⟩ ↔ |f⟩ transition with an additional pump tone at frequency of |g⟩ ↔ |f⟩ transition is applied. **d**, Spectroscopy measurements of the |g⟩ ↔ |f⟩ transition.

## Phononic Crystal

The phonon crystal supports numerous resonant modes, a subset of which exhibit weak (in value) interactions with the |g⟩↔|e⟩ transition of the artificial atom, visible as smaller avoided crossings in Fig. S2b. Our analysis focuses exclusively on the fundamental mode, which possesses the highest quality factor and exhibits the strongest coupling to both the artificial atom and the input/output IDTs of the crystal. We characterize this fundamental mode by measuring its transmission spectrum. The data is fitted using a standard resonator transmission formula: $|S_{21}(f)|^2 = a^2 \times \frac{Q_l^2}{1+4Q_l^2\left(\frac{f}{f_0}-1\right)^2}$ where $Q_l$ is loaded

quality factor of the resonator, $f_0$ is a resonator frequency, f is a frequency of the probe tone and $a$ is a normalization factor that incorporates the external quality factor and the attenuation of the microwave line. By performing this fit at different probe powers, we observe the power dependence of the quality factor. It increases from $Q_0 = 24 \times 10^3$ for low number of phonons to $Q_m = 34 \times 10^3$ at high phonon number. From the fit, we extract normalization factor of a = $10^{-5}$. Using a reference transmission level (see Measurement setup) we estimate the external quality factor for the output port of the acoustic resonator $Q_{out} = 2.4 \times 10^6$.

## Avoided crossing modification and amplification at three-level atom pumping

In Fig. S3a transmission through a resonator is shown in a wide range of magnetic flux bias, when transition $|g\rangle \leftrightarrow |f\rangle$ is pumped. The energy splitting $E_{eg}$ of $|g\rangle \leftrightarrow |e\rangle$ transition reaches minimum at the centre of the plots similarly to Fig. S2a. In resonance between the atom and the resonator, two symmetric anticrossing splitting are observed. The picture is strongly modified, when the pumping power is increased and at power higher than -18 dBm transmission becomes more than one, indicating amplification.

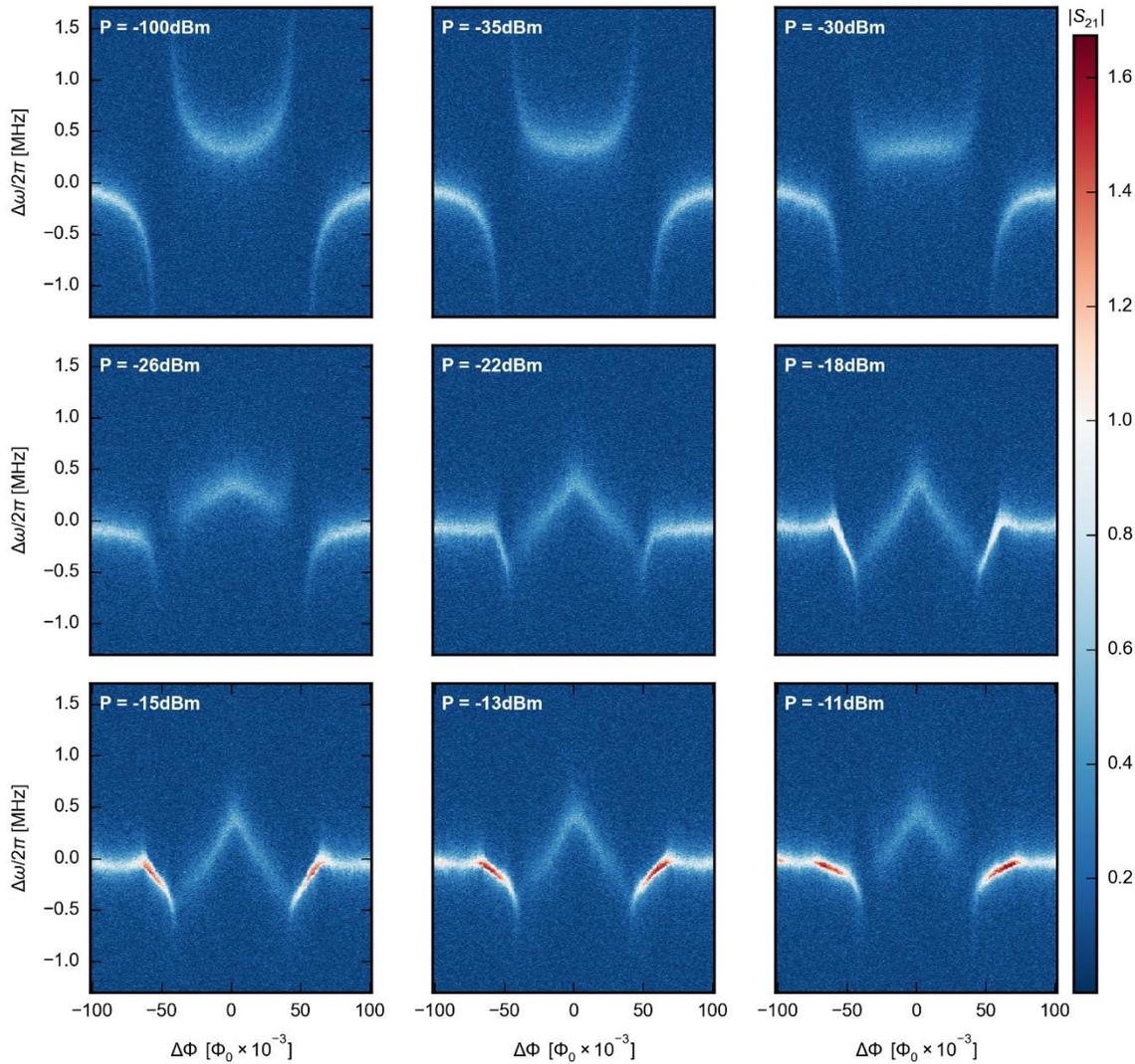

**Figure S3 |** Transmission through the resonator versus flux bias through the atom. Pumping power at $|g\rangle \rightarrow |f\rangle$ transition is

increased from the upper left to the down right.

## Emission versus SASER parameters

We studied the emission power dependence on pump power of the artificial atom. Figure S4a shows emission spectra of the SASER from pump power of the atom. Extracted values of the phonon number and FWHM for the data of Fig. S4a are displayed on Fig. S4b, where we can see clear threshold dependence, optimal regime of the SASER and critical pump power. The experimental data is accompanied by analytical calculations of the simplified SASER model with parameters extracted from the experimental data (Fig. S4c,d).

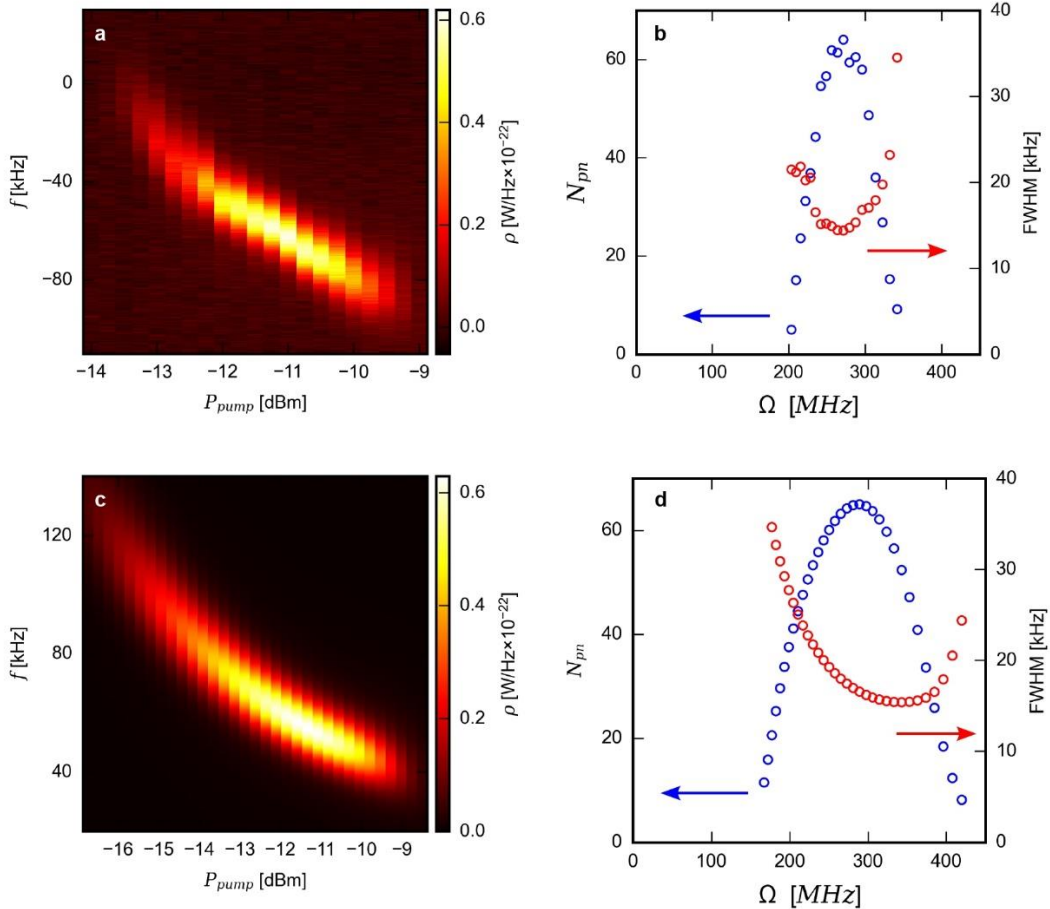

**Figure S4** | **a**, Power spectroscopy of the incoming emission from the resonator at the different pump power of the artificial atom. **b**, Estimated phonon number and FWHM for each spectrum from Fig. S4a. **c**, Simulation of the SASER operation Fig. S4a with parameters, extracted from the experiment. **d**, Phonon number and FWHM from the simulation Fig. S4c.